\documentclass[a4paper,12pt]{article}
\usepackage{graphicx,amssymb,bm,latexsym,color}
\newcommand{\be}{\begin{equation}}
\newcommand{\ee}{\end{equation}}
\newcommand{\bea}{\begin{eqnarray}}
\newcommand{\eea}{\end{eqnarray}}
\newcommand{\ena}{\end{eqnarray}}

\usepackage{graphicx,amssymb,amsmath,bm,latexsym}

\makeatletter
\@addtoreset{equation}{section}

\makeatother
\newcommand{\vs}[1]{\vspace{#1 mm}}

\renewcommand{\t}{\theta}

\newcommand{\pa}{\partial}
\newcommand{\nn}{\nonumber\\}
\newcommand{\p}[1]{(\ref{#1})}

\begin{document}

\begin{titlepage}

\begin{flushright}
KU-TP 043 \\
\end{flushright}

\vskip .5in

\begin{center}

{\Large\bf Dynamical solutions in the 3-Form Field Background in
the Nishino-Salam-Sezgin Model}
\vskip .5in

{\large
Masato Minamitsuji,$^{a,b,}
$\footnote{E-mail address:masato.minamitsuji"at"kwansei.ac.jp}
Nobuyoshi Ohta$^{c,}$\footnote{E-mail address: ohtan"at"phys.kindai.ac.jp}
and
Kunihito Uzawa$^{c,}$
}\\
\vs{10}
$^a${\em Center for Quantum Spacetime, Sogang University,
Shinsu-dong 1, Mapo-gu, Seoul, 121-742 South Korea} \\
$^b${\em Department of Physics, Graduate School of Science and Technology,
Kwansei Gakuin University, Sanda 669-1337, Japan}\\
$^c${\em Department of Physics, Kinki University,
Higashi-Osaka, Osaka 577-8502, Japan} \\

\begin{abstract}
We investigate the dynamical 3-form flux compactifications
and their implications for cosmology in our brane world
in the six-dimensional Nishino-Salam-Sezgin model,
which is usually referred to as the Salam-Sezgin model.
We take the background of the 3-form field acting on the internal space
and timelike dimensions without the $U(1)$ gauge field strength.
The first class of solutions we discuss is
the dynamical generalization of the static solutions obtained recently.
In this class, we find that the time evolution is restricted only in
the azimuthal dimension of the internal space and not in the ordinary
three-dimensional one, which does not give a cosmological evolution.
The second class of solutions is obtained by exchanging the roles of the radial
coordinate and the time coordinate from the assumptions in the first class.
At the center of the internal space, there is a conical singularity
which may be interpreted as our 3-brane world,
but it is difficult to realize a warped structure in the
direction of the ordinary 3-space.
Except for the oscillating solutions, a dynamical
evolution leads to expanding or contracting ordinary 3-space,
depending on the choice of time-direction.
Furthermore, in the expanding solutions, there are
decelerating and accelerating ones:
In the latter solution, the evolution is sustained
for a finite proper time and the scale factor of the 3-space diverges.

\end{abstract}

\end{center}

\end{titlepage}

\newpage
\setcounter{page}{1}
\section{Introduction}

Brane world models have attracted particular interests in recent years.
In the cosmological aspects, six-dimensional brane world
models may be useful for a resolution of the cosmological
constant problem~\cite{cc+} (see, however,~\cite{cc-}).
A six-dimensional model is also recognized as an important
playground to study cosmology and gravity
with stabilized extra dimensions by fluxes
of antisymmetric tensor fields.
In the simplest realization of the flux compactifications in six dimensions,
the internal space has the shape of a rugby ball~\cite{cc+,cc-},
where codimension-two branes are placed at the poles.
The warped generalizations of the rugby ball solutions
were reported in the Nishino-Salam-Sezgin (NSS) six-dimensional
supergravity~\cite{ggp}
and in the Einstein-Maxwell theory~\cite{msyk}.
See~\cite{ns} for the NSS model.
Note that this model is usually referred to as the Salam-Sezgin
model especially when compactification is discussed.
However, the compactification on $S^2$, which was found by Salam and Sezgin,
was constructed in the theory first proposed by Nishino and Sezgin.
Therefore, it is more suitable for us to
refer to the model as the 
NSS model.
In Ref.~\cite{cdgv}, a study of the cosmology was presented
on the analogy of the classical mechanics.
It is also recognized that a 3-brane in six or higher dimensions
faces a problem about the localization of the matter,
because such a brane generically produces a curvature singularity.
A way to circumvent this problem
is to introduce a thickness of the brane, as proposed in~\cite{pst,vc}.

In order to explore the cosmology, one useful approach is to use
the dynamical solutions.
In this direction, exact dynamical solutions were obtained
in~\cite{6dt,cs,km,shock}.
The field components of the bosonic part of the NSS model
are the gravity, the $U(1)$-gauge field,
Kalb-Ramond 2-form (i.e., 3-form field strength) and the dilaton.
In Refs.~\cite{6dt,cs}, warped time dependent solutions with
the $U(1)$-field strength,
without the Kalb-Ramond 3-form field strength
in the NSS model, were explored.
The solutions have all six dimensions evolving in time.
There are three types of the exact
time dependent solutions:
One of them is the scaling solution,
in which both the scale factors of the external and internal spaces
have the linear dependence on the proper time.
The other two solutions have more complicated time dependence.
In the earlier times, these solutions
can be seen as the generalizations of the solutions
obtained in Ref.~\cite{mn},
but in the later times both approach the above scaling solution.
The scaling solution is the generalization of the static solution
obtained in \cite{ggp} and becomes an attractor.
In this sense, the static brane solution
in the 2-form background is always unstable.
An extension of the time dependent 2-form solution in
the NSS model to more general Einstein-Maxwell-dilaton models
was considered in Ref.~\cite{km}.
The exact wave solution was also obtained in~\cite{shock}.

The purpose of this work is to present the other classes of
time dependent solutions in the NSS model in the presence of
the nonvanishing 3-form field strength
[but in the absence of the $U(1)$ field strength],
and discuss their implications for the brane world cosmology.
The first class of solutions is the dynamical generalization of
the static solutions obtained recently in Ref.~\cite{ap},
where the internal space has a torus topology with two cusps,
along the line of the similar generalization of brane solutions~\cite{MOU}.
We  will find that our solutions are stable, in contrast to
the 2-form solutions.
The second class is obtained by exchanging the roles of the radial
coordinate and the time coordinate in the first class.
We also discuss the dynamical generalization of
the black 1-brane solution in the NSS model.

The paper is constructed as follows.
In Sec.~2, we briefly review the NSS model.
In Sec.~3, the time dependent generalization of the solutions
and their application in constructing the brane world model are discussed.
In Sec.~4, the S-brane-like solution,
obtained by exchanging the roles of space and time,
and their cosmological properties are discussed.
The final section is devoted to give a brief summary
and conclusion.
In the Appendix, we discuss the black 1-brane solution.

\section{Nishino-Salam-Sezgin model}

The action of the bosonic part of the NSS
supergravity~\cite{ns}
is given by
\bea
I = \int d^6 x \sqrt{-g} \left[ \frac{1}{2\kappa^2}
( R - (\pa \phi)^2 ) - \frac{1}{4} e^{-\phi} F_{\mu\nu}^2
-\frac1{12} e^{-2\phi}H_{\mu\nu\rho}^2-\frac{2g^2}{\kappa^4}e^{\phi} \right],
\label{action}
\ena
where $\phi$ is the dilaton, $H=dB$
is the field strength for the Kalb-Ramond field $B$
and $F=dA$ is the field strength for the $U(1)$ gauge field $A$.
The parameters $g$ and $\kappa$ are
gauge and gravitational coupling constants, respectively.
Variation of the action (\ref{action}) gives the
field equations:
\bea
R_{\mu\nu} = \pa_\mu \phi \pa_\nu \phi
+ \kappa^2 e^{-\phi}\Big(F_{\mu\nu}^2-\frac18 F^2 g_{\mu\nu}\Big)
+\frac{\kappa^2}2 e^{-2\phi}\Big(\Big(H^2\Big)_{\mu\nu}-\frac16 H^2 g_{\mu\nu}\Big)
+\frac{g^2}{\kappa^2}e^\phi g_{\mu\nu}.
\label{Einstein}
\ena
\bea
\Box \phi +\frac{\kappa^2}{4} e^{-\phi}F^2+\frac{\kappa^2}{6}e^{-2\phi}H^2
-\frac{2g^2}{\kappa^2} e^\phi=0,
\label{dila}
\ena
\bea
\pa_{\mu} \left(
\sqrt{- g} e^{-2 \phi} H^{\mu\nu \rho} \right) = 0,
\label{field}
\ena
\bea
\pa_{\mu} \left(\sqrt{- g} e^{- \phi} F^{\mu\nu} \right) = 0.
\ena
Most of the preceding works have discussed
the $U(1)$ field acting on the internal space dimensions,
$F\neq 0$,
with the vanishing 3-form $H=0$.
Instead, in this work,
we consider the nonvanishing 3-form field
acting on the internal space and timelike dimensions
$H\neq 0$, with the vanishing $U(1)$ field $F=0$.

\section{Generalized warped compactifications by the 3-form}

\subsection{Ansatz and equations}

In this section, we take the following metric ansatz~\cite{MOU}:
\bea
ds_6^2 = - e^{2 u_0(t,y)} dt^2 + e^{2 u_1(t,y)}\sum_{i=1}^{3} (dx^i)^2
+e^{2v(t,y)} dy^2 +e^{2w(t,y)}d\t^2,
\label{met1}
\ena
where the coordinates $t$ and $x^i$ describe our four-dimensional worldvolume,
and the remaining $y$ and $\t$ are transverse to it.
The metric components $u_0, u_1, v, w$ and the dilaton $\phi$
are assumed to be functions of both $t$ and $y$.
For convenience of our derivation,
we also define
\bea
U \equiv u_0+3u_1-v+w.
\label{u}
\ena
We assume that
for the 3-form field strength there is only the $y$ dependence and also
the $U(1)$ field strength vanishes:
\bea
H = 
E'(t,y)
\, dt \wedge dy \wedge d\t, \qquad
F=0.
\label{eleb}
\ena
Throughout the main text in this paper,
the dot and prime denote derivatives with respect to $t$
and $y$.
The Einstein equations~\p{Einstein} become
\bea
\label{E1}
&& e^{2(u_0-v)}(u_0''+ U' u_0')-\ddot U + \ddot u_0 - 2\ddot v
+\dot u_0(\dot U -\dot u_0+2\dot v)-3\dot u_1^2-\dot v^2-\dot w^2 \nn
&& \qquad = \dot \phi^2+\frac{\kappa^2}{2}e^{-2\phi-2v-2w} {E'}^2
-\frac{g^2}{\kappa^2}e^{\phi+2u_0}, \\
\label{E2}
&& -3\dot u_1'-\dot w'+u_0'(3\dot u_1+\dot w)-3 u_1'(\dot u_1-\dot v)+w'(\dot v-\dot w)
=\dot \phi \phi', \\
\label{E3}
&& e^{2(u_1-v)}(u_1''+ U' u_1')-e^{-2(u_0-u_1)}[ \ddot u_1
+\dot u_1(\dot U -2 \dot u_0+2\dot v)] \nn
&& \qquad =
-\frac{\kappa^2}{2}e^{-2\phi-2v-2w} {E'}^2
-\frac{g^2}{\kappa^2}e^{\phi+2u_0}, \\
\label{E4}
&&U''+v''+{u_0'}^2+3{u_1'}^2+{w'}^2-v'(U'+v')-e^{-2(u_0-v)}[\ddot v
+\dot v(\dot U-2\dot u_0+2\dot v)] \nn
&& \qquad = -{\phi'}^2+\frac{\kappa^2}{2}e^{-2\phi-2u_0-2w} {E'}^2
-\frac{g^2}{\kappa^2} e^{\phi+2v},\\
\label{E5}
&& e^{2(w-v)}(w''+ U' w')-e^{2(w-u_0)}[ \ddot w +\dot w(\dot U -2 \dot u_0+2\dot v)] \nn
&& \qquad = \frac{\kappa^2}{2}e^{-2\phi-2u_0-2v} {E'}^2 -\frac{g^2}{\kappa^2}e^{\phi+2w}.
\ena
The dilaton Eq.~\p{dila} and the equation for the
3-form field~\p{field} give
\bea
\label{dil}
&& (e^U \phi')'-(e^{U-2u_0+2v}\dot\phi)^{\bm \cdot}
-\kappa^2 e^{-u_0+3u_1-v-w-2\phi}{E'}^2
-\frac{2g^2}{\kappa^2}e^\phi=0, \\
\label{fs1}
&& (e^{U-2u_0-2w-2\phi}E')'=0, \\
\label{fs2}
&& (e^{U-2u_0-2w-2\phi}E')^{\bm \cdot}=0.
\ena
The latter Eqs.~\p{fs1} and \p{fs2} give
\bea
\label{int1}
e^{U-2u_0-2w-2\phi}E'=c,
\ena
where $c$ is an integration constant.
Substituting \p{int1} into Eqs.~\p{E1}, \p{E3}, \p{E5} and \p{dil},
we obtain
\bea
\label{E11}
\Big( e^U u_0'-\frac{\kappa^2}{2}c E \Big)'
&=& e^{U+2(v-u_0)}[\ddot U - \ddot u_0 +2\ddot v
-\dot u_0(\dot U -\dot u_0+2\dot v)+3\dot u_1^2+\dot v^2+\dot w^2+\dot\phi^2] \nn
&& -\frac{g^2}{\kappa^2}e^{\phi+2v+U},\\
\label{E21}
\Big( e^U u_1'+\frac{\kappa^2}{2}c E \Big)'
&=& (\dot u_1 e^{U+2(v-u_0)})^{\bm \cdot} -\frac{g^2}{\kappa^2}e^{\phi+2v+U},\\
\label{E51}
\Big( e^U w'-\frac{\kappa^2}{2}c E \Big)'
&=& (\dot w e^{U+2(v-u_0)})^{\bm \cdot} -\frac{g^2}{\kappa^2}e^{\phi+2v+U}, \\
\label{dil1}
\Big(e^U \phi' -\kappa^2 c E \Big)'
&=& (\dot \phi e^{U+2(v-u_0)})^{\bm \cdot} +\frac{2g^2}{\kappa^2}e^{\phi+2v+U}.
\ena
Let us assume that the solutions satisfy the following equations:
\bea
\label{E12}
\Big( e^U u_0'-\frac{\kappa^2}{2}c E \Big)'
&=& -\frac{g^2}{\kappa^2}e^{\phi+2v+U} +\ell_0,\\
\label{E22}
\Big( e^U u_1'+\frac{\kappa^2}{2}c E \Big)'
&=& -\frac{g^2}{\kappa^2}e^{\phi+2v+U}+\ell_1,\\
\label{E52}
\Big( e^U w'-\frac{\kappa^2}{2}c E \Big)'
&=& -\frac{g^2}{\kappa^2}e^{\phi+2v+U}+\ell_w, \\
\label{dil2}
\Big(e^U \phi' -\kappa^2 c E \Big)'
&=& \frac{2g^2}{\kappa^2}e^{\phi+2v+U}-2\ell_\phi,
\ena
where $\ell_i$ ($i=0,1,w,\phi$)
are constants corresponding to the time derivative parts
in Eqs.~\p{E11}--\p{dil1}.
Relaxing these separability conditions
is a very interesting issue, but we find it difficult to do so.
As in Ref.~\cite{MOU}, we also assume that $U$ is independent of $y$.
In fact, it is known that some nontrivial time dependent solutions can be
obtained even with this restriction.
Then using $v'=u_0'+3u_1'+w'$ obtained from $U'=0$
and Eqs.~\p{E12}--\p{E52}, we find
\bea
(2v+\phi)''=-\frac{8g^2}{\kappa^2} e^{2v+\phi}+2(\ell_0+3\ell_1+\ell_w-\ell_\phi)e^{-U}.
\label{solphi}
\ena
It is hard to find a solution unless we assume
\bea
\ell_0+3\ell_1+\ell_w-\ell_{\phi}=0.
\label{cons}
\ena
When this is obeyed, the solution to Eq.~\p{solphi} is
\bea
e^{-(2v+\phi)} = \frac{4g^2}{\kappa^2 f_1^2} \cosh^2[f_1(y-y_1)],
\label{makiko}
\ena
where $f_1$ and $y_1$ are constants.
Although relaxing the condition Eq. (\ref{cons})
would be also an interesting issue,
we leave it for future study.

It follows from Eqs.~\p{E12}--\p{dil2} that
\bea
\label{E13}
u_0' &=& \frac{\kappa^2}{2}c e^{-U}E
-\frac{f_1}{4} \tanh[f_1(y-y_1)] -G_0 + \ell_0 e^{-U} y \,,
\\
\label{E23}
u_1' &=& -\frac{\kappa^2}{2}c e^{-U}E-\frac{f_1}{4}
\tanh[f_1(y-y_1)] -G_{1} + \ell_1 e^{-U} y \,,
 \\
\label{E53}
w' &=& \frac{\kappa^2}{2}c e^{-U}E-\frac{f_1}{4}
\tanh[f_1(y-y_1)] -G_w +\ell_w e^{-U} y ,\\
\label{dil3}
\phi' &=& \kappa^2 c e^{-U}E+\frac{f_1}{2}
\tanh[f_1(y-y_1)] +2G_{\phi} - 2 \ell_\phi e^{-U} y \,,
\ena
where we have chosen $G_i$ to be constant for simplicity.
We then find from \p{makiko} and \p{dil3}
\bea
\label{v3}
v' = -\frac{\kappa^2}{2}c e^{-U}E-\frac{5f_1}{4}
\tanh[f_1(y-y_1)] -G_{\phi} + \ell_\phi e^{-U} y .
\ena
The condition $U'=0$ imposes
\bea
G_0+3G_1+G_w-G_{\phi}=0.
\eea
For convenience, we define
\bea
Y \equiv 2\kappa^2 c e^{-U}E.
\ena
By substituting the above relations into Eq.~\p{E4},
we find
\bea
0&=&\frac{1}{2}\big(Y^2-Y'\big)
-f_1^2+e^{-U}\Big(\ell_{\phi}-\ell_v\Big)
\nonumber\\
&+&3 A_1^2+A_w^2+3A_\phi^2+(3A_1+A_w-A_\phi)^2+Y(3A_1+A_\phi),
\label{E41}
\ena
where we have assumed $\big(\dot{v} e^{U-2u_0+2v}\big)^{\cdot}=\ell_v$
($\ell_v$ is a constant),
and have defined
\bea
A_i=G_i-\ell_i e^{-U}y,~~(i=1,w,\phi).
\ena
At this stage, except for the cases where $\ell_i=0$,
it is still not easy to find analytically tractable solutions.
Thus, in this paper, we focus on the case $\ell_i=0$,
leaving study of more general cases for the future.
Then, without loss of generality, we may set $3G_1+G_{\phi}=0$.
Now Eq.~\p{E41} reduces to
\bea
Y^2-Y'-f_2^2=0,
\label{sol2}
\ena
where we have defined
\bea
f_2^2 = 2f_1^2 - 4(G_\phi-G_w)^2-\frac{32}{3} G_\phi^2.
\label{shouryuu}
\ena
$f_2^2$ can take either positive, zero and negative value
and the solutions for  $Y(y)$ can be classified into
the three types, depending on the sign of $f_2^2$.
We will give the solutions shortly.
Once $Y(y)$ is obtained, by integrating Eqs.~\p{E13}--\p{v3},
we can find the metric components and dilaton as
\bea
\label{E14}
u_0 &=& -\frac{1}{4}
  P(y)
 -\frac{1}{4} \ln\cosh[f_1(y-y_1)]
-(2G_\phi-G_w)y +h_0(t), \\
\label{E24}
u_1 &=& \frac{1}{4}
 P(y)
-\frac{1}{4} \ln\cosh[f_1(y-y_1)]
+ \frac13 G_\phi y +h_1(t), \\
\label{v4}
v &=& \frac{1}{4}
P(y)
-\frac{5}{4} \ln\cosh[f_1(y-y_1)]
-G_\phi y +h_v(t), \\
\label{E54}
w &=& -\frac{1}{4}
 P(y)
 -\frac{1}{4} \ln\cosh[f_1(y-y_1)]
- G_w y +h_w(t), \\
\label{dil4}
\phi &=& -\frac{1}{2}
 P(y)
+\frac{1}{2} \ln\cosh[f_1(y-y_1)]
+ 2G_\phi y +h_\phi(t),
\ena
where the function $P(y)$ depends on the sign of $f_2^2$;
hence, the type of solutions,
and $h_i$ ($i=0,1,v,w,\phi$) could be functions of $t$.
When all $h_i$ vanish, the solution becomes static.

Now the three types of solutions are given as follows:
\paragraph{The sinh solutions:}
For a positive $f_2^2$,
the solution to Eq.~\p{sol2} is given by
\bea
Y(y)=-f_2 \coth[f_2(y-y_2)].
\ena
After integration, we obtain
\bea
P(y)=\ln \big|\sinh[f_2(y-y_2)]\big|.
\eea

\paragraph{The sin solutions:}
For a negative $f_2^2$,
the solution to Eq.~\p{sol2} is given by
\bea
Y(y)=-|f_2| \cot[|f_2|(y-y_2)].
\eea
After integration, we obtain
\bea
\label{E14c}
P(y)=\ln \big|\sin[|f_2|(y-y_2)]\big|.
\eea

\paragraph{The linear solutions:}
For $f_2^2=0$, the solution to Eq.~\p{sol2} is given by
\bea
Y(y)=-\frac{1}{y-y_2}.
\eea
After integration, we obtain
\bea
\label{E14d}
P(y)=\ln \big|y-y_2|.
\eea

These give the dynamical generalization of the static solutions found in~\cite{ap}.
The correspondence is given by the following reparameterizations
of $f_1\leftrightarrow \lambda_2$, $f_2\leftrightarrow \lambda_1$ and
$q\leftrightarrow c$.
Note, however, that Eq. (\ref{shouryuu}) does not
completely agree with Eq. (2.10) of~\cite{ap},
in some factors.

\subsection{General time dependent solutions}

Now our task is to solve the remaining time derivative parts of (\ref{E21})-(\ref{dil1})
and \eqref{E4}.
The remaining time dependent parts of these equations give
\bea
\ddot h_i =(\dot h_0
-3\dot h_1-\dot h_v-\dot h_w) \dot h_i,~~
\ena
where $i=1,v,w, \phi$.
They lead to
\bea
\dot h_i e^{-h_0+3h_1+h_v+h_w} &=& c_i,
\label{1st}
\ena
where all $c_i$ are constants.
The off-diagonal components of the Einstein equation~\p{E2} imposes
the constraints among the time dependent functions
\bea
\dot h_v=-3 \dot h_1, \quad
\dot h_\phi = 6 \dot h_1, \quad
\big(11G_\phi-3G_w\big)\dot h_1 = \big(G_w-G_\phi\big) \dot h_w.
\label{2nd}
\ena

It is straightforward to confirm that
except for the case of $G_{w}= G_{\phi}$
all $h_i$ cannot have any nontrivial time dependence.
To show this, for the moment let us set $h_0(t)=0$,
which is always possible by an appropriate rescaling
for the time coordinate.
{}From Eq.~(\ref{2nd}), we have $c_v=-3c_1$ and $c_\phi=6c_1$.

In the case of $11G_{\phi}\neq 3G_w$,
Eq.~(\ref{1st}) tells us that
\bea
&&\dot h_1=\frac{G_w-G_{\phi}}{(11G_{\phi}-3G_w)t},\quad
\dot h_v=-3\frac{G_w-G_{\phi}}{(11G_{\phi}-3G_w)t},
\nonumber\\
&&
\dot h_w=\frac{1}{t},\quad
\dot h_v=6\frac{G_w-G_{\phi}}{(11G_{\phi}-3G_w)t}.
\eea
It follows from Eq.~(\ref{E11}) that
\bea
0=\frac{48(G_\phi-G_w)^2}{(11G_{\phi}-3G_w)^2 t^2}\,.
\eea
Thus, the consistency requires $G_{w}=G_{\phi}$.
In the case of $11G_{\phi}=3G_w$, $\dot h_w=0$.
We may set $h_w=0$ and then $3h_1+h_v=$ const.
{}From Eq.~(\ref{1st}), we have $h_1= C t$, $h_v= -3 Ct$ and
$h_{\phi}=6C t$, where $C$ is a constant.
We then find from Eq.~(\ref{E11})
\bea
C^2=0\,,
\eea
which leads to a trivial solution.
Thus, $G_{w}=G_\phi$ must be satisfied to have nontrivial solutions.

For $G_w=G_{\phi}$,
\bea
f_2^2=2f_1^2-\frac{32}{3}G_w^2.
\eea
Then, ${\dot h}_{v}=\dot h_{\phi}=\dot h_1=0$
but
$\dot h_{w}\neq 0$.
Thus we obtain
\bea
h_1=k_1,\quad
h_v=k_v,\quad
h_{\phi}=k_{\phi},
\ena
where $k_1$, $k_v$ and $k_{\phi}$ are integration constants.
Therefore, the ordinary 3-space metric and the dilaton do not depend on time.
{}From Eq. (\ref{makiko}), we obtain
\bea
k_{\phi}=-2
\left(k_v+\ln\Big(\frac{2g}{\kappa f_1}\Big)
\right).
\eea
$h_0$ and $h_w$ are still undetermined except for
the relation \p{1st}, which gives
$(e^{h_w})^{\cdot}=c_w e^{k_v+3k_1} e^{h_0}$.
The two-dimensional metric in the $t$ and $\theta$ directions
now becomes
\bea
ds^2&=&
e^{2F(y)}
\big(-e^{2h_0}dt^2+e^{2h_w}d\theta^2\big)
\nonumber \\
&=&
\frac{e^{2F(y)}}{c_w^2 e^{2(k_v+3k_1)}}
\left(
-\Big(\big(e^{h_w}\big)^{\cdot}dt\Big)^2
+c_w^2 e^{2(k_v+3k_1)}e^{2h_w}d\theta^2
\right)
\nonumber\\
&=&
\frac{e^{2F(y)}}{c_w^2 e^{2(k_v+3k_1)}}
\left(
-dT^2
+c_w^2 e^{2(k_v+3k_1)}T^2 d\theta^2
\right),
\eea
where $F(y)$ is the common $y$ dependent part
and $T=e^{h_w(t)}$ is the proper time.
Thus, the time dependence only appears in the $\theta$ direction
and not in the ordinary 3 directions, which cannot present
any cosmological evolution.
These results imply that the solution is dynamically unstable for
the evolution along the $\theta$ direction, but stable for the others.
In terms of the four-dimensional effective theory, the effective
potential would contain one flat direction associated with the motion
in the $\theta$ direction.

One might guess that more generalized time dependent
solutions could be obtained by allowing
for the time variations of $f_i$ and $G_j$.
However, in such cases, the time dependence of the Einstein and dilaton equations
cannot be separated out from the dependence on $y$.
Thus, it seems to be impossible to find consistent time dependent solutions.

\subsection{Comparison with 2-form flux compactification }

It would be interesting to compare our result with
the dynamical solutions with the 2-form background in the NSS model.
In the case of the $U(1)$ background,
there is the so-called scaling solution \cite{6dt,cs},
which can be the attractor.
In this solution, both the internal and external dimensions
have the time dependence linear in time and evolve uniformly.
It means that the static solutions in the $U(1)$ background
are not stable. Thus,
the scaling symmetry always requires some additional mechanism
to stabilize the extra dimensions, for example, as discussed in
\cite{abpq,bht}.

In our 3-form background,
the analyses in the previous subsection indicate that
there may be no time dependent generalizations of
the static solutions in the case of $G_{w}\neq G_{\phi}$.
For the case $G_{w}=G_\phi$,
the time evolution is allowed only in the $\theta$ direction
and is forbidden into the other directions.
Thus in either case, this seems to imply that the 3-form compactification
is stable in comparison with the case of $U(1)$-background.
Here one remark is in order. In deriving the solutions,
we made some assumptions on the time dependence
and relaxation of them might lead to additional possible evolutions of space.
At the moment, we do not have the complete proof on the stability of such
general solutions.

\subsection{Application to brane world model}

\subsubsection{Singularity}

Reference~\cite{ap} investigated the construction of the brane world
by using the static solution in the background of the 3-form field strength.
A curvature singularity exists at $y=y_2$, irrespective of the
type of solutions.
Note that this is a real curvature singularity, which is more severe
than a conical one and may not be identified as
our 3-brane world.
To see this, we define $\bar y:=y-y_2$, and
the approximate spacetime metric near the singularity $\bar y=0$
is given by
\bea
ds^2\sim  c_y^2{\bar y}^{1/2} d\bar y^2
+c_{\theta}^2 \frac{d\theta^2}{\bar y^{1/2}}
-c_t^2 \frac{dt^2}{\bar y^{1/2}}
+c_x^2 {\bar y}^{1/2}\sum_{i=1}^{3} (dx^i)^2\,,
\eea
where $c_i$ ($i=y,\theta,t,x$) are unimportant coefficients.
This metric leads to a divergent scalar curvature $R\sim {\bar y}^{-5/2}$
near $\bar y=0$. Thus, as we will discuss in the next subsection,
to construct a brane world, the singularity is first removed from the original
spacetime in~\cite{ap}.
Then, an identical copy of the remaining piece of the spacetime
is attached to the opposite side
across the codimension-one boundary at $y=y_2+\epsilon$, where $\epsilon>0$,
which wraps the axis of the rotational symmetry of the internal space.
This boundary may be identified as our brane world.
There are jumps of the physical quantities across the boundary.

\subsubsection{Construction of the brane world}

We now construct the brane world by employing the dynamical solution
in the 3-form background.
First, the original spacetime is cut at $y=y_0>y_1(>y_2)$
and the singular part of $y_2<y<y_0$ is removed.
Then, an identical copy of the remaining piece is glued
with the original one, at the codimension-one boundary $y=y_0$.

The induced metric on the boundary is given by
\bea
ds_5^2=e^{2w(t,y_0)}d\theta^2-e^{2u_0(t,y_0)}dt^2+e^{2u_1(t,y_0)}\sum_{i=1}^{3} (dx^i)^2.
\eea
The extrinsic curvature to the boundary is given by
\bea
K_{ab}= \frac{\epsilon}{2e^{v}} \big(g_{ab}\big)'\Big|_{y_0},
\eea
where $\epsilon=+ 1$ or $-1$ denotes the direction of
the normal vector toward increasing or decreasing $y$,
respectively.
In our case, we choose $\epsilon=+1$.
Now the extrinsic curvature is discontinuous across the boundary
and its jump is determined by the Israel junction condition:
\bea
\Big[g_{ab}\Big]=0,\quad
\Big[{\bar K}_{ab}\Big]
=\Big[K_{ab}-g_{ab}K\Big]
=-\kappa^2 S_{ab},
\eea
where $[A]$ denotes the jump of a physical quantity $A$
across $y=y_0$, and $S_{ab}$ is
the stress-energy tensor of the matter localized on the boundary.
This codimension-one boundary can be identified as our brane world.

For our background metric, we obtain
\bea
&&{\bar K}{}^{\theta}{}_{\theta}
=-
e^{-v}\big(u_0'+3u_1'\big),
\quad
{\bar K}{}^{t}{}_{t}
=-
e^{-v}\big(w'+3u_1'\big),
\quad
\frac{1}{3}{\bar K}{}^{i}{}_{i}
=-
e^{-v}\big(u_0'+w'+2u_1'\big).
\nonumber\\
&&
\eea
$S_{ab}$ also can be decomposed into the components
$S^{t}_{t}=-\rho$,
$S^{\theta}_{\theta}=p_{\theta}$
and $(1/3)S^{i}_{i}=p$.
The junction condition at $y=y_0$
is given by
\bea
&&
\Big(e^{-v}\big(
u_0'+3u_1'
\big)\Big)|_{y=y_0}
=\frac{\kappa^2}{2}p_{\theta},
\label{jk1-1}
\\
&&
\Big(e^{-v}\big(
w'+3u_1'
\big)\Big)|_{y=y_0}
=-\frac{\kappa^2}{2}\rho,
\label{jk2-1}
\\
&&
\Big(e^{-v}\big(
w'+u_0'+2u_1'
\big)\Big)|_{y=y_0}
=\frac{\kappa^2}{2}p.
\label{jk3}
\eea
They can be rewritten as
\bea
&&
\Big(e^{-v}\big(
u_1'-w'
\big)\Big)|_{y=y_0}
=\frac{\kappa^2}{2}
\big(p_{\theta}-p\big),
\label{jk4}
\\
&&
\Big(e^{-v}\big(
u_1'-u_0'
\big)\Big)|_{y=y_0}
=-\frac{\kappa^2}{2}\big(\rho+p\big),
\\
\label{jk5}
&&
\Big(e^{-v}\big(
u_0'+3u_1'
\big)\Big)|_{y=y_0}
=\frac{\kappa^2}{2}p_{\theta}.
\label{jk6}
\eea

Our discussion in this subsection can be applied
to the {\it sinh} and {\it linear} solutions
discussed in the previous section,
and the form of $P(y)$
in Eqs. \eqref{E14}-\eqref{dil4} is not specified.
It is now clear that the zero-thickness limit of $y_0\to y_2$
is not well behaved, since the left-hand side of the junction Eqs.
\eqref{jk4}-\eqref{jk6} becomes singular.
Note that the $sin$ solution contains another curvature singularity
at $y=\pi/|f_2|+y_2$ and is excluded from our consideration.
In addition, for the $sinh$ or {\it linear} solutions,
the finiteness of bulk volume is not ensured.
Thus, following Ref.~\cite{ap}, here
we put the second brane world at some $y= L>y_0$.
Note that the second brane takes $\epsilon=-1$.
The junction condition at $y=L$ is given in a similar way
with the opposite sign of the right-hand side of Eqs. (\ref{jk1-1})-(\ref{jk3}).

In order to obtain a dynamical solution, we must choose $G_w=G_\phi$.
Also, we may choose the integration constants as
\bea
k_1=0,\quad k_v=0,\quad k_{\phi}
=-2\ln\Big(\frac{2g}{\kappa f_1}\Big).
\eea
and thus $\big(e^{h_w}\big)^{\cdot}=c_w^{-1} e^{h_0}$.
By introducing the proper coordinate $T=c_w e^{h_w}$,
after some computations, we obtain
\bea
&&u_1-w
=\frac{1}{2}
 P(y)
+\frac{4}{3}G_{\phi}y-\ln T,
\\
&&
u_1-u_0
=\frac{1}{2} P(y)
+\frac{4}{3}G_{\phi}y,
\label{u1u0}
\\
&&3u_1+u_0
=\frac{1}{2}
 P(y)
- \ln \cosh \big[f_1(y-y_1)\big].
\eea
The junction conditions are reduced to
\bea
&&
\label{jk1}
\Big\{
e^{-v}\Big(
\frac{1}{2}P'(y_0)
+\frac{4}{3}G_{\phi}\Big)
\Big\}
= \frac{\kappa^2}{2}\big(p_{\theta}-p\big),
\\
&&
\label{jk2}
\Big\{
e^{-v}\Big(
\frac{1}{2}P'(y_0)
+\frac{4}{3}G_{\phi}\Big)
\Big\}
= -\frac{\kappa^2}{2}\big(p+\rho\big),
\\
&&
 \Big\{
e^{-v}\Big(
P'(y_0)
-f_1\tanh \big[f_1(y_0-y_1)\big]
\Big)
\Big\}
=\frac{\kappa^2}{2}p_{\theta}.
\eea
Noting that $v$ is not an explicit function of $T$,
the left-hand side of the junction equations
become static,
and $\rho$, $p$ and $p_{\theta}$ remain constant.
In the static case without the condition $G_w=G_{\phi}$,
generically the three quantities
$\rho$, $p$ and $p_{\theta}$ satisfy the different
equations of state from the tension.
Now, however, from Eq. (\ref{jk1}) and (\ref{jk2}),
the inclusion of the time dependence, i.e., the condition $G_w=G_{\phi}$,
induces the relation $\rho=-p_{\theta}$.

Furthermore, the requirement of the recovery of the Lorentz symmetry at
the brane $y=y_0$, i.e., $u_1'=u_0'$, together with Eq.~(\ref{u1u0})
leads to
\bea
3P'(y_0)
=-8G_{\phi}.
\eea
By combining this with the previous relations,
$u_0'=u_1'=w'$ is obtained.
This relation suggests that $\rho=-p=-p_{\theta}$.
Thus, we are lead to the conclusion that
the brane with the ordinary four-dimensional spacetime with
the Lorentz symmetry is supported only by the tension.
Note that this conclusion is derived with some assumptions
concerning time dependence,
and without them we might be able to construct
the brane world with the Lorentz symmetry,
supported by some matter other than the tension.

Before closing this section, following Ref. \cite{ap},
the four-dimensional effective theory on the brane world
should be discussed.
In the standard warped solutions, the four-dimensional
Planck mass is obtained by integrating over the two-internal space.
However, in our case, in general, the warp function in the timelike direction
is different from that of the ordinary 3-space.
Then,
the six-dimensional Einstein-Hilbert term now reduces to
\bea
&&M_6^4 \int d^6 x \sqrt{-g}R
\nonumber\\
&\sim&
2(2\pi)M_6^4
\int d^4 x\sqrt{-g^{(4)}}
\Big(
 \int_{y_0}^L dy e^{-u_0+3u_1+v+w} g^{(4)tt}R^{(4)}_{tt}
+\int_{y_0}^L dy e^{u_0+u_1+v+w} g^{(4)ij}R^{(4)}_{ij}
\Big),
\nonumber\\
&&
\eea
where $M_6:=\big(\frac{1}{2\kappa^2}\big)^{1/4}$ is the six-dimensional
Planck mass.
Note that the factor 2 in front of the integration
come from the $Z_2$-symmetry, and $(2\pi)$
is from the integration over the azimuthal direction.
To define the unique four-dimensional effective Planck mass $M_p$,
we need to require that
\bea
 2(2\pi)M_6^4  \int_{y_0}^L dy e^{-u_0+3u_1+v+w}
=2(2\pi)M_6^4\int_{y_0}^L dy e^{u_0+u_1+v+w}
=:M_p^2,
\eea
or more explicitly
\bea
\int_{y_0}^L dy\frac{P(y)e^{2(G_{\phi}-G_w)y}}
               {\cosh^2\big[f_1(y-y_1)\big]}
=\int_{y_0}^L dy\frac{e^{-\frac{8}{3}G_{\phi}y}}
               {\cosh^2\big[f_1(y-y_1)\big]}.
\eea
In the case of $G_{\phi}\neq G_w$
there is no time dependent solution.
The cutoff parameter $y_0$ appears in the definition of $M_p^2$.
In other words,
the effect of the brane thickness is renormalized into the four-dimensional
Planck mass.
In the case of $G_{\phi}=G_w$, $e^{w}\propto T$
and hence $M_p^2\propto T$.
In this case, we should move to the Einstein frame.
Then, the conformal transformation to the Einstein frame,
$g^{(E)}_{\mu\nu}= T g^{(4)}_{\mu\nu}$,
leads to the cosmic expansion of $\tau^{1/3}$,
where $\tau$ is the cosmic proper time defined in the Einstein frame.
However, this power in the expansion law
is not sufficient for obtaining realistic cosmology.
Realistic cosmological evolutions may be obtained
by considering the time dependent matter on the brane,
through the induced motion of the brane in the bulk.

\section{S-brane-like solutions}

\subsection{Field equation}

Assuming the same metric ansatz~(\ref{met1}) as in the previous section,
we investigate another class of the time dependent solutions
by exchanging the roles of the $t$-coordinate
with the $y$ coordinate
as well as those of $u_0$ with $v$.
The way of constructions of solutions are very similar to
the case of S-branes, see e.g., \cite{sb1,sb2,sb4,sb5}.
We also assume that the 3-form field strength
is the function of time
\bea
H = \dot E(t,y) \, dt \wedge dy \wedge d\t .
\label{elec}
\ena
The equations of motion for the flux are given by
\bea
\label{fs12}
&& (e^{U-2u_0-2w-2\phi}\dot E)'=0, \\
\label{fs22}
&& (e^{U-2u_0-2w-2\phi}\dot E)^{\bm \cdot}=0.
\ena
Equations~\p{fs12} and \p{fs22} give
\bea
\label{int12}
e^{\Phi-2v-2w-2\phi}\dot{E}=c,
\ena
where $c$ is a constant and
$$\Phi :=U+2v-2u_0.$$
In this section, we assume $\Phi=0$ is independent of $t$,
$\dot \Phi=0$.
Employing \p{int12}, the Einstein and dilaton equations are now given by
\bea
\label{E112}
&&\left(e^{U}u_0{}' \right)'
= e^{\Phi}\big(\ddot u_0
-\dot u_0^2+3\dot u_1^2+\dot v^2+\dot w^2+\dot\phi^2\big)
+\frac{1}{2}\kappa^2 c\dot E -\frac{g^2}{\kappa^2}e^{\phi+2v+U},\\
\label{E212}
&&\left(e^{U}u_1{}' \right)'
=(\dot u_1 e^{\Phi} -\frac{1}{2}\kappa^2 c E )^{\bm \cdot}
 -\frac{g^2}{\kappa^2}e^{\phi+2v+U},
\\
&&e^{U}
\big(v''+U''-v'{}^2-v'U'+3u_1{}'{}^2+w'{}^2+u_0{}'{}^2 +\phi'{}^2\big)
\nonumber \\
&&= (\dot v e^{\Phi} +\frac{1}{2}\kappa^2 c E
)^{\bm \cdot} -\frac{g^2}{\kappa^2}e^{\phi+2v+U},
\\
\label{E512}
&&\left(e^{U}w{}' \right)'
=(\dot w e^{\Phi} +\frac{1}{2}\kappa^2 c E
)^{\bm \cdot} -\frac{g^2}{\kappa^2}e^{\phi+2v+U},
\\
\label{dil12}
&&\left(e^{U}\phi{}' \right)'
= (\dot \phi  e^{\Phi} +\kappa^2 c E )^{\bm \cdot} +\frac{2g^2}{\kappa^2}e^{\phi+2v+U}.
\ena
Let us look for the solutions with the following assumptions that
\bea
\label{E112_2}
&&
0= e^{\Phi}\big(\ddot u_0
-\dot u_0^2+3\dot u_1^2+\dot v^2+\dot w^2+\dot\phi^2\big)
+\frac{1}{2}\kappa^2 c\dot E -\frac{g^2}{\kappa^2}e^{\phi+2v+U},\\
\label{E212_2}
&&
0=(\dot u_1 e^{\Phi} -\frac{1}{2}\kappa^2 c E )^{\bm \cdot}
 -\frac{g^2}{\kappa^2}e^{\phi+2v+U},
\\
&&0= (\dot v e^{\Phi} +\frac{1}{2}\kappa^2 c E
)^{\bm \cdot} -\frac{g^2}{\kappa^2}e^{\phi+2v+U},
\\
\label{E512_2}
&&0
=(\dot w e^{\Phi} +\frac{1}{2}\kappa^2 c E
)^{\bm \cdot} -\frac{g^2}{\kappa^2}e^{\phi+2v+U},
\\
\label{dil12_2}
&&
0= (\dot \phi  e^{\Phi} +\kappa^2 c E )^{\bm \cdot} +\frac{2g^2}{\kappa^2}e^{\phi+2v+U}.
\ena
By combining Eq. (\ref{E212_2})-(\ref{dil12_2}),
with $\dot \Phi=0$ and hence $\dot u_0=\dot v+3\dot u_1+\dot w$,
we find
\bea
(2u_0+\phi)^{\bm \cdot\bm\cdot}=\frac{8g^2}{\kappa^2} e^{2u_0+\phi}.
\label{flash}
\ena
For Eq. (\ref{flash}), there are three classes of solutions, namely
the {\it sinh}, {\it sin} and {\it linear} solutions.
We discuss these solutions in order.

\paragraph{The sinh solutions:}
A solution to Eq. (\ref{flash}) is given by
\bea
e^{-(2u_0+\phi)} = \frac{4g^2}{\kappa^2 f_1^2} \sinh^2[f_1(t-t_1)],
\ena
where $f_1$ and $t_1$ are constants.
Assuming that the right-hand sides of
Eqs.~\eqref{E112}--\eqref{dil12} vanish,
we obtain
\bea
\label{E152}
\dot v &=&
- \frac{\kappa^2}{2}c  e^{-\Phi}E-\frac{f_1}{4}
\coth[f_1(t-t_1)]
-G_v,
 \\
\label{E232}
\dot u_1 &=&
 \frac{\kappa^2}{2}c e^{-\Phi} E-\frac{f_1}{4}
\coth[f_1(t-t_1)]
-G_1,\\
\label{E532}
\dot w &=&
-\frac{\kappa^2}{2}c  e^{-\Phi}E-\frac{f_1}{4}
\coth[f_1(t-t_1)]
-G_w,
\\
\label{dil32}
\dot \phi &=&- \kappa^2 c  e^{-\Phi} E+\frac{f_1}{2}\coth[f_1(t-t_1)]
+2G_{\phi},
\ena
where $G_i$ ($i=v,1,w,\phi$) are also constants.
Because of the assumption that $\dot \Phi=0$, we obtain
\bea
\label{v322}
\dot u_0
=\frac{\kappa^2}{2}c e^{-\Phi} E-\frac{5f_1}{4}
\coth[f_1(t-t_1)]
-G_{\phi},
\ena
and
\bea
3G_1-G_{\phi}+G_v+G_w=0.
\eea
Substituting these into Eq.~\p{E112},
we find
\bea
&& \frac12 (Y^2+\dot Y)
-(3G_1+G_{\phi})Y
\nn&&
-f_1^2
+\big(-3G_1+G_{\phi}-G_w\big)^2
+3G_1^2+G_w^2+3G_\phi^2=0,
\label{E412}
\ena
where $Y$ is defined as in the previous section:
\bea
Y \equiv 2\kappa^2 c E e^{-\Phi}.
\ena
Without loss of generality,
we may impose the further condition that
\bea
3G_1+G_{\phi}=0.
\ena
Then Eq.~\p{E412} reduces to
\bea
Y^2+\dot Y+f_2^2=0,
\label{sol22}
\ena
where we defined
\bea
f_2^2 = -2f_1^2 + 4(G_\phi-G_w)^2+\frac{32}{3} G_\phi^2.
\ena
Note that $f_2^2$ can be either positive, negative or zero.

For a positive, negative and vanishing $f_2^2$,
the solution to Eq.~\p{sol22} is given by
\bea
&&Y=f_2 \cot[f_2(t-t_2)], \quad |f_2| \coth[|f_2|(t-t_2)],\quad
\frac{1}{t-t_2},
\ena
respectively.
The corresponding solutions with the possible $y$ dependence
are given by
\bea
u_0 &=& \frac{1}{4} Q(t)
-\frac{5}{4} \ln \Big|\sinh[f_1(t-t_1)]\Big|
-G_\phi t +h_0(y), \\
u_1 &=& \frac{1}{4}Q(t)
 -\frac{1}{4} \ln\Big|\sinh[f_1(t-t_1)]\Big|
+ \frac{1}{3} G_\phi t+h_1(y), \\
v &=& -\frac{1}{4}Q(t)
-\frac{1}{4} \ln\Big|\sinh[f_1(t-t_1)]\Big|
-\big(2G_\phi-G_w\big) t +h_v(y)
, \\
w &=& -\frac{1}{4}Q(t)
-\frac{1}{4} \ln\Big|\sinh[f_1(t-t_1)]\Big|
- G_w t+h_w(y)
, \\
\phi &=& -\frac{1}{2}Q(t)
 +\frac{1}{2}\ln\Big|\sinh[f_1(t-t_1)]\Big|
+ 2G_\phi t+h_\phi(y),
\ena
where for a positive, negative or vanishing $f_2^2$,
$Q(t)$ is defined by
\bea
Q(t):=\ln \Big|\sin[f_2(t-t_2)]\Big|,\quad
\ln \Big|\sinh[|f_2|(t-t_2)]\Big|,\quad
\ln |t-t_2|,
\eea
respectively.


\paragraph{The sin solutions:}
Another solution to Eq. (\ref{flash}) is the oscillatory type:
\bea
e^{-(2u_0+\phi)} = \frac{4g^2}{\kappa^2 f_1^2} \sin^2[f_1(t-t_1)].
\ena
Then, defining the new parameter
\bea
f_2^2 = 2f_1^2 + 4(G_\phi-G_w)^2+\frac{32}{3} G_\phi^2,
\ena
which is always positive for a positive $f_1^2$,
we get the corresponding solutions with the possible $y$ dependence:
\bea
u_0 &=& \frac{1}{4}\ln \Big|\sin[f_2(t-t_2)]\Big|
-\frac{5}{4} \ln\Big|\sin[f_1(t-t_1)]\Big|
-G_\phi t +h_0(y), \\
u_1 &=& \frac{1}{4}\ln \Big|\sin[f_2(t-t_2)]\Big|
 -\frac{1}{4} \ln\Big|\sin[f_1(t-t_1)]\Big|
+ \frac{1}{3} G_\phi t+h_1(y), \\
v &=& -\frac{1}{4}\ln \Big|\sin[f_2(t-t_2)]\Big|
-\frac{1}{4} \ln\Big|\sin[f_1(t-t_1)]\Big|
-\big(2G_\phi-G_w\big) t +h_v(y)
, \\
w &=& -\frac{1}{4}\ln \Big|\sin[f_2(t-t_2)]\Big|
-\frac{1}{4} \ln\Big|\sin[f_1(t-t_1)]\Big|
- G_w t+h_w(y)
, \\
\phi &=& -\frac{1}{2}\ln \Big|\sin[f_2(t-t_2)]\Big|
 +\frac{1}{2}\ln\Big|\sin[f_1(t-t_1)]\Big|
+ 2G_\phi t+h_\phi(y).
\ena

\paragraph{The linear solutions:}
The last solution to Eq. (\ref{flash}) takes the form
\bea
e^{-(2u_0+\phi)} = \frac{4g^2}{\kappa^2} (t-t_1)^2.
\ena
Then, by defining
\bea
f_2^2 =  4(G_\phi-G_w)^2+\frac{32}{3} G_\phi^2,
\ena
which is always positive, we get
the corresponding solutions with the possible $y$ dependence:
\bea
u_0 &=& \frac{1}{4}\ln \Big|\sin[f_2(t-t_2)]\Big|
-\frac{5}{4} \ln |t-t_1|
-G_\phi t +h_0(y), \\
u_1 &=& \frac{1}{4}\ln \Big|\sin[f_2(t-t_2)]\Big|
 -\frac{1}{4} \ln |t-t_1|
+ \frac{1}{3} G_\phi t+h_1(y), \\
v &=& -\frac{1}{4}\ln \Big|\sin[f_2(t-t_2)]\Big|
-\frac{1}{4} \ln |t-t_1|
-\big(2G_\phi-G_w\big) t +h_v(y)
, \\
w &=& -\frac{1}{4}\ln \Big|\sin[f_2(t-t_2)]\Big|
-\frac{1}{4} \ln |t-t_1|
- G_w t+h_w(y)
, \\
\phi &=& -\frac{1}{2}\ln \Big|\sin[f_2(t-t_2)]\Big|
 +\frac{1}{2}\ln |t-t_1|
+ 2G_\phi t+h_\phi(y),
\ena
where
\bea
f_2^2 = 2f_1^2 + 4(G_\phi-G_w)^2+\frac{32}{3} G_\phi^2.
\ena

In all the three types of solutions,
setting all $h_i=0$ leads to purely time dependent solutions.

\subsection{Generalization to the $y$ dependent cases}

Following the similar arguments as done in the previous section,
we find that the $y$ dependence can be included only
for the choice of $G_{\phi}=G_w$.
In any case of the $sinh$, $sin$ and {\it linear} solutions
shown in the previous subsection,
the $y$ dependent functions satisfy
\bea
&&h_0=k_0,\quad h_{\phi}=k_{\phi},\quad h_1=k_1,
\quad
\left(e^{h_w}\right)'
=c_w e^{-k_0-3k_1}e^{h_v},
\label{fuge}
\eea
where $k_{0}$, $k_{v}$ and $k_{1}$ are integration constants
satisfying
\bea
k_{\phi}+2k_0
=-2\ln\left(\frac{2g}{\kappa f_1}\right)\,.
\eea
Thus, even taking into account the $y$ dependence,
the spacetime structure in the ordinary four dimensions
does not contain a warped structure.
The internal space metric can be written as
\bea
ds_2^2&=&
 F(t)^2\big(e^{2h_v}dy^2+e^{2h_w}d\theta^2\big)
\nonumber\\
&=&\frac{e^{2(k_0+3k_1)} F(t)^2}
        {c_w^2}
\left(
\left(\big(e^{h_w}\big)' dy\right)^2
+\frac{c_w^2}{e^{2(k_0+3k_1)}}
\big(e^{h_w}\big)^2 d\theta^2
\right)
\nonumber\\
&=&\frac{e^{2(k_0+3k_1)} F(t)^2}
        {c_w^2}
\left(
dR^2
+\frac{c_w^2}{e^{2(k_0+3k_1)}}
R^2 d\theta^2
\right)\,,
\eea
where $R=e^{h_w}$ is the proper radial coordinate.
Assuming the standard periodicity $2\pi$
for the angular coordinate,
there is generically a conical singularity at the center
\bea
\label{deficit}
\Delta=2\pi \left(1-\frac{c_w}{e^{k_0+3k_1}} \right).
\eea
This conical singularity at $R=0$ can be seen as our brane world.
The ordinary 3-space metric cannot depend on the internal space coordinate and
therefore it is impossible to realize a warped structure.

\subsection{Cosmological behaviors on the brane}

Let us now discuss the cosmological behaviors of our solutions.
In our real Universe, the scale factor is not oscillating.
Thus, in this subsection
we do not consider the solutions with oscillating scale factor and
focus only on the $sinh$ solutions of $G_{w}=G_{\phi}$
with $f_2^2\leq 0$.
Without loss of generality,
we may assume $f_1>0$.
For simplicity, we discuss the behaviors in the large $t$ limit:
\bea
&&
u_0\approx \frac{1}{4}
\left(\sqrt{2\Big(1-\frac{16}{3}g_{\phi}^2\Big)}-5-4g_{\phi}\right)f_1t
=:p_0 f_1t,
\nonumber\\
&&
u_1\approx \frac{1}{4}
\left(\sqrt{2\Big(1-\frac{16}{3}g_{\phi}^2\Big)}-1
+\frac{4}{3}g_{\phi}\right)f_1t=:p_1 f_1t,
\nonumber\\
&&
v=w\approx -\frac{1}{4}
\left(\sqrt{2\Big(1-\frac{16}{3}g_{\phi}^2\Big)}+1
+4g_{\phi}\right)f_1t
=:p_v f_1t,
\eea
where we also define $g_{\phi}:=G_{\phi}/f_{1}$.
Note that $|g_{\phi}|\leq \sqrt{3}/4\approx 0.433$.
For $-0.196<g_{\phi}<0.410$, $p_1>0$
and for $g_{\phi}>-0.395$, $p_v<0$ (see Fig. 1 and 2).
\begin{figure}
\begin{minipage}[t]{.45\textwidth}
\label{fig2}
   \begin{center}
    \includegraphics[scale=.60]{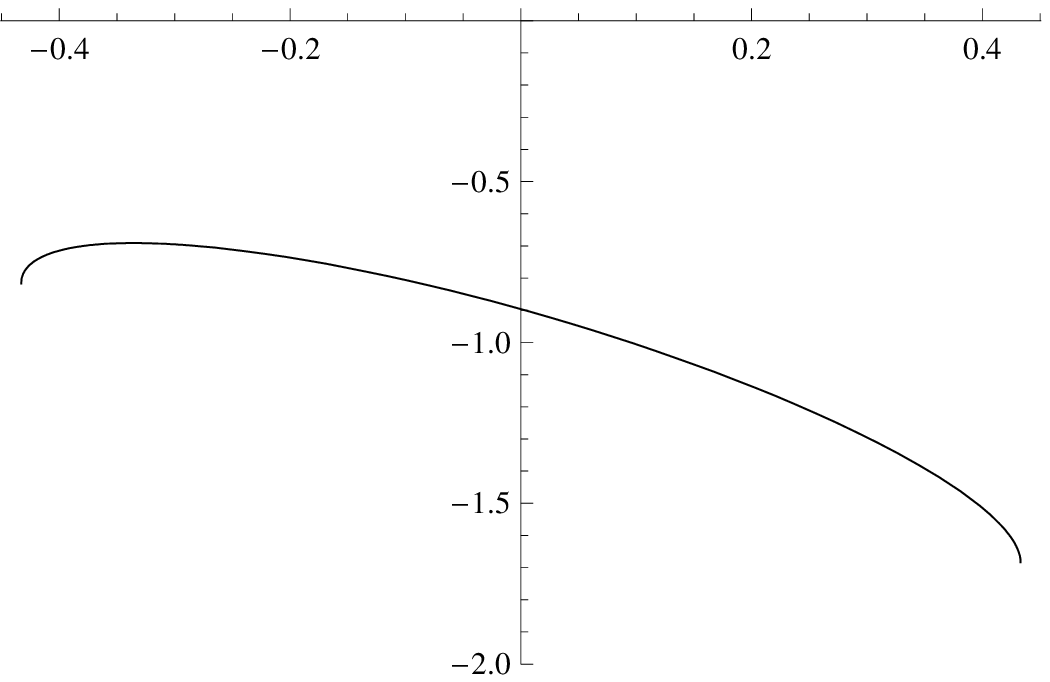}
        \caption{
The plot for $p_0$ is shown as a function of $g_{\phi}$ for $f_1>0$.
}
   \end{center}
 \end{minipage}
\hspace{0.3cm}
\begin{minipage}[t]{.45\textwidth}
\label{fig3}
   \begin{center}
    \includegraphics[scale=.60]{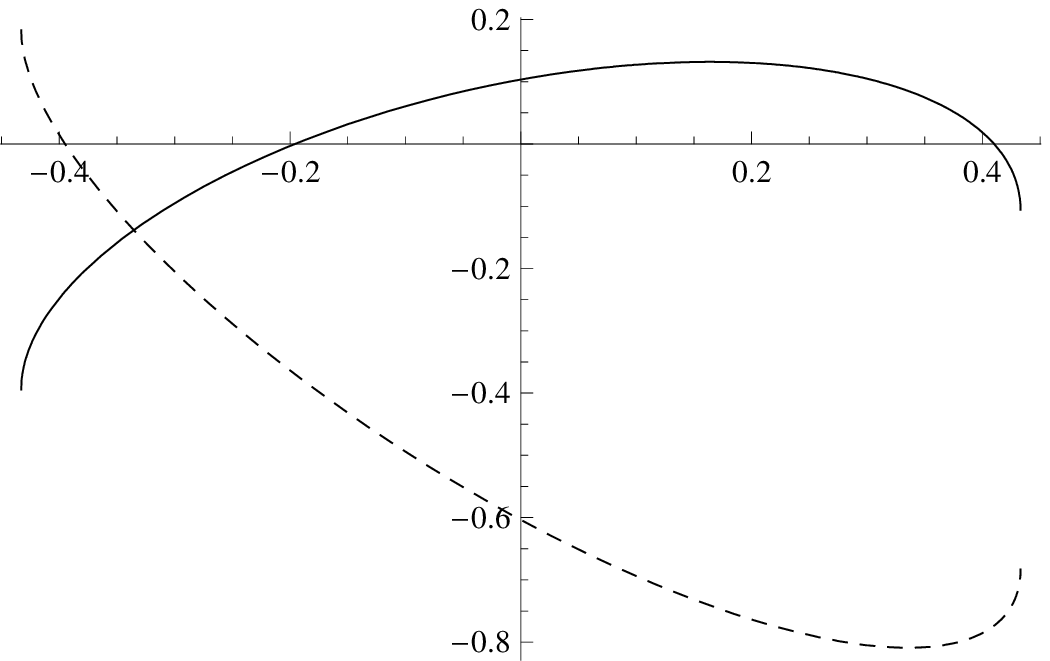}
        \caption{
The plots for $p_1$ (the solid curve) and $p_v$ (the dashed curve) are shown
as a function of $g_{\phi}$ for $f_1>0$.
}
   \end{center}
 \end{minipage}
\end{figure}

In one choice of the proper time coordinate, $dT=-e^{u_0}dt$,
the approximate spacetime metric is given by
\bea
\label{asymp_1}
ds^2= \big(-T\big)^{2q_v}
\Big(dR^2+R^2\big(1-\frac{\Delta}{2\pi}\big)^2d\theta^2\Big)
-dT^2
+\big(-T\big)^{2q_1}\sum_{i=1}^{3} (dx^i)^2,
\eea
where unimportant constants are eliminated by the proper rescalings of
$R$ and $x^i$
and
$\Delta$ is defined in Eq. (\ref{deficit}).
We also defined the powers
\bea
q_1:=-\frac{p_1}{|p_0|},\quad
q_v:=-\frac{p_v}{|p_0|}.
\eea
In the other choice of the proper time coordinate, $dT=e^{u_0}dt$,
the approximate spacetime metric becomes
\bea
\label{asymp_2}
ds^2= T^{2q_v}
\Big(dR^2+R^2\big(1-\frac{\Delta}{2\pi}\big)^2d\theta^2\Big)
-dT^2
+T^{2q_1}\sum_{i=1}^{3} (dx^i)^2.
\eea
In Fig 3, we show the behavior of $q_1$ and $q_v$.
For $-0.196<g_{\phi}<0.410$, $q_1<0$.
For $g_{\phi}>-0.395$, $q_v>0$.
For $-\sqrt{3}/4<g_{\phi}<-0.335$, $q_1>q_v$, while
for the rest $q_1<q_v$.
\begin{figure}
\begin{minipage}[t]{.45\textwidth}
\label{fig4}
   \begin{center}
    \includegraphics[scale=.60]{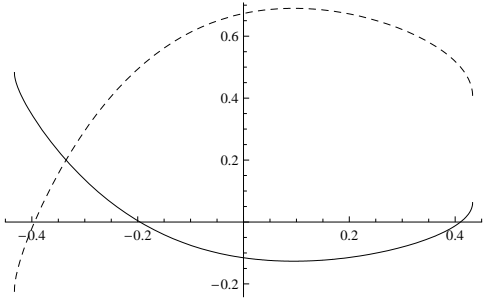}
        \caption{
The plots for $q_1$ (the solid curve)
and $q_v$ (the dashed curve)
are shown as a function of $g_{\phi}$ for $f_1>0$.
}
\end{center}
\end{minipage}
\end{figure}

For $-0.196<g_{\phi}<0.410$, $q_1<0$ and $q_v>0$.
Thus, in the metric (\ref{asymp_1}),
in the $T\to 0-$ limit,
the size of the ordinary 3-space increases and diverges
within a finite time,
while that of the internal space is shrinking to zero.

For $-\sqrt{3}/4<g_{\phi}<-0.335$, $q_1>q_v$ and $q_1>0$.
Thus, for this range of $g_{\phi}$,
the metric (\ref{asymp_2})
describes the ordinary 3-space which is
expanding faster than the internal space dimensions.
In particular, for $g_{\phi}<-0.395$,
the internal space is contracting.
But since $q_1\leq 0.406$,
an accelerating expansion, as in the inflationary or dark energy Universe,
cannot happen.

\paragraph{The Einstein frame:}
Let us briefly discuss how our solution behaves in the Einstein frame.
Rewriting the original metric in this form,
\bea
ds^2=g^{(4)}_{\mu\nu}dx^{\mu}dx^{\nu}+e^{2v}dy^2+e^{2w}d\theta^2,
\eea
we obtain
\bea
R=R^{(4)}+2\dot{v}^2+2\dot{w}^2+\cdots,
\eea
where $R^{(4)}$ is the Ricci scalar associated with the metric
$g^{(4)}_{\mu\nu}$.
The gravity action can reduce to
\bea
\int d^6 x\sqrt{-G}R
&\sim& V_2 \int d^4 x \sqrt{-g_{(4)}}e^{v+w}R^{(4)}
\sim V_2 \int d^4 x \sqrt{-g_{(E)}}R^{(E)}
\eea
where $V_2=\int dv dw$ is the comoving volume of the internal space,
which is assumed to be finite for instance by introducing a cutoff.
The Einstein frame metric is given by
\bea
 ds_E^2:=
g^{(E)}_{\mu\nu}dx^{\mu}dx^{\nu}=e^{v+w}g^{(4)}_{\mu\nu}dx^{\mu}dx^{\nu}
=-e^{v+w+2u_0}dt^2+e^{v+w+2u_1}\sum_{i=1}^{3} (dx^i)^2.
\eea
Using the solutions, we find
\bea
v+w+2u_1=\frac{1}{3}\big(v+w+2u_0\big)
=-\ln\Big|\sinh[f_1(t-t_1)]\Big|-\frac{4}{3}G_\phi t.
\eea
Thus, in the proper time coordinate system $d\tau=\pm e^{v+w+2u_0}dt$,
the Einstein frame metric can be written as
\bea
ds_E^2\approx -d\tau^2+\big(\mp \tau\big)^{2/3}\sum_{i=1}^{3} (dx^i)^2,
\eea
which corresponds to a contracting or expanding
Universe filled by the stiff matter.


\section{Conclusions}

We have investigated the dynamical 3-form flux compactifications
and their implications for brane world cosmology
in the six-dimensional Nishino-Salam-Sezgin model.
We take the background of the 3-form field acting on the internal space
and timelike dimensions without the $U(1)$ gauge field strength.

The first class of solutions we discussed was
the dynamical generalization~\cite{MOU} of the static solutions,
recently obtained in \cite{ap}. It turned out that
the dynamical generalization is possible only for the special case.
In this class, we found that the time evolution is restricted in
the azimuthal dimension of the internal space.
The ordinary three-dimensional space is not dynamical and hence does not
give rise to a cosmological evolution.
There is a curvature singularity at the boundary of the internal space,
which is more severe than the conical one.
To construct the brane world model, following \cite{ap}, we first
removed the singular part from the original solution
and then glued the remaining piece of the spacetime
with an identical copy at the codimension-one boundary
wrapping the axis of the rotational symmetry.
This boundary may be regarded as our 3-brane world.
Generically, because of the presence of the 3-form field
the Lorentz symmetry in the ordinary four-dimensional spacetime is broken.
In general, such a brane can be supported
by the matter with the energy density $\rho$,
the pressure in the ordinary 3-space $p$
and that in the azimuthal dimension $p_{\theta}$,
which would be different from the pure tension.
Since the inclusion of the time dependence reduces the number of the parameters,the brane can be embedded by the matter with $p_{\theta}=-\rho$
(but still $p\neq -\rho$).
However, at the place where the Lorentz symmetry is recovered,
the boundary brane world can be supported only by the pure tension, i.e.,
$p=p_{\theta}=-\rho$.
Note the solutions reported here and related implications for
the brane world model
were obtained under several assumptions.
It is very interesting to explore the time dependent solutions
and associated brane world models without these assumptions.

The second class of models was obtained by exchanging the roles of the radial
coordinate and the time coordinate from the first class one.
At the center of the internal space, there is a conical singularity
which may be interpreted as codimension-two our 3-brane world.
But it does not seem to be possible to realize a warped structure in the external
dimensions, as expected in analogy with the previous case.
Except for the oscillating dynamical solutions,
the solutions led to the expanding or contracting ordinary 3-space,
depending on the choice of time direction.
Among the expanding solutions, there are
decelerating and accelerating ones:
In the latter solution, the scale factor in the ordinary 3-space diverges
within a finite time. In the Einstein frame, however,
the Universe always followed the expansion law
of the one filled by the stiff matter, irrespective of the
choice of parameters.

\section*{Acknowledgements}

The work of M.M was supported by the National Research Foundation of Korea(NRF)
grant funded by the Korea government(MEST) under Grant No. 20090063070.
The work of N.O. was supported in part by the Grant-in-Aid for
Scientific Research Fund of the JSPS under Grant
Nos. 20540283 and  21$\cdot$\,09225.
K.U. would like to thank 
H. Kodama, 
M. Sasaki and 
T. Okamura for continuing encouragement. K.U. is supported
by the Grant-in-Aid for Young Scientists (B) of JSPS Research, under
Contract No. 20740147.
Part of this work was carried out while N.O. was visiting ITP, Beijing.
He thanks 
R.G. Cai for discussions and kind hospitality.


\appendix

\section{Dynamical black 1-branes }

In this Appendix, we discuss another class of the dynamical solutions
with the nonvanishing 3-form field strength.
The solution represents the black 1-brane extended along the $y$ direction
in the presence of a cosmological constant.
We now take the following ansatz for the metric
\bea
ds_6^2 = e^{2u(t,y,r)}\big(-dt^2 + dy^2\big)
+e^{2v(t,y,r)}(dr^2+r^2 d\Omega_3^2),
\ena
the dilaton $\phi=\phi(t,y,r)$ and
the 3-form field strength
\bea
H=E(t,y,r){}_{,r}dt\wedge dr\wedge dy.
\ena
The $(t,y)$ coordinates cover the worldvolume of the 1-brane, while
the rest does the transverse spatial dimensions.
The equations for the 3-form field are given by
\bea
\label{MOU8}
&& (e^{-2u+2v-2\phi} r^3 E_{,r})^{\bm \cdot}
=(e^{-2u+2v-2\phi} r^3 E_{,r})_{,y}
=(e^{-2u+2v-2\phi} r^3 E_{,r})_{,r}=0.
\ena
Defining
$
V \equiv 2u+2v,
$
the integration of Eq. \p{MOU8} gives
\bea
r^3 e^{V-4u-2\phi}E_{,r}=c,
\ena
where $c$ is an integration constant.
Then, the diagonal components of Einstein equations
and the dilaton equation of motion
are given by
\bea
\label{MOU12}
&&e^{-2u}
\big(-u_{,yy}-4u_{,y}v_{,y}
-\ddot{u}+2\ddot{v}+\ddot{V}
-2\dot{u}\dot{v}
-\dot{u}\dot{V}
+2\dot{u}^2
+4\dot{v}^2
\big)
+e^{-2v}\big(-u_{,r}V_{,r}-u_{,rr}-\frac{3}{r}u_{,r}\big)
\nonumber\\
&&=-e^{-2u}\dot{\phi}^2-\frac{\kappa^2 c}{2r^3}e^{-V-2v}E_{,r}
 +\frac{g^2}{\kappa^2}e^{\phi}
,\\
\label{MOU22}
&&
e^{-2u}
\big(u_{,yy}-2v_{,yy}-V_{,yy}
+2u_{,y}v_{,y}+u_{,y}V_{,y}
-2u_{,y}^2-4v_{,y}^2
+\ddot{u}
+\dot{u}\big(\dot{V}-2\dot{v}+2\dot{u}\big)
\big)
\nonumber\\
&&+e^{-2v}\big(-u_{,r}V_{,r}-u_{,rr}-\frac{3}{r}u_{,r}\big)
=e^{-2u}\phi_{,y}^2-\frac{\kappa^2 c}{2r^3}e^{-V-2v}E_{,r}
 +\frac{g^2}{\kappa^2}e^{\phi},
\\
\label{MOU32}
&&e^{-2u}\big(\ddot{v}+\dot{v}(\dot{V}-2\dot{u}+2\dot{v})
-v_{,yy}-v_{,y}(V_{,y}-2u_{,y}+2v_{,y})
\big)
\nonumber \\
&&+e^{-2v}
 \big(-V_{,rr}-v_{,rr}-\frac{3}{r}v_{,r}
 +v_{,r}V_{,r}-2(u_{,r}^2+v_{,r}^2)
\big)
=e^{-2v}\phi_{,r}{}^2 -\frac{\kappa^2 c}{2r^3}e^{-V-2v}E_{,r}
 +\frac{g^2}{\kappa^2}e^{\phi},
\\
\label{MOU42}
&&e^{-2u}\big(\ddot{v}+\dot{v}(\dot{V}-2\dot{u}+2\dot{v})
-v_{,yy}-v_{,y}(V_{,y}-2u_{,y}+2v_{,y})
\big)
\nonumber\\
&&+e^{-2v}
\big(-v_{,rr}-\frac{3}{r}v_{,r}-\big(v_{,r}+\frac{1}{r}\big)V_{,r}\big)
= \frac{\kappa^2 c}{2r^3}e^{-V-2v}E_{,r}
 +\frac{g^2}{\kappa^2}e^{\phi},\\
\label{MOU52}
&&
 e^{-2u}\big(-\ddot{\phi}-(\dot{V}-2\dot{u}+2\dot{v})\dot{\phi}
+\phi_{,yy}+(V_{,y}-2u_{,y}+2v_{,y})\phi_{,y}\big)
\nonumber \\
&&+e^{-2v}\big(\phi_{,rr}+\big(V_{,r}+\frac{3}{r}\big)\phi_{,r}\big)
-\frac{\kappa^2 c}{r^3}e^{-V-2v}E_{,r}
-\frac{2g^2}{\kappa^2}e^{\phi}
=0.
\eea
We require in Eq. (\ref{MOU12}) and (\ref{MOU22})
\bea
u_{,r} = \frac{\kappa^2}{2r^3}ce^{-V}E.
\label{s1}
\ena
Similarly in Eqs. \p{MOU42} and \p{MOU52},
\bea
v_{,r}= -\frac{\kappa^2}{2r^3}ce^{-V}E, \quad
\phi_{,r} = \frac{\kappa^2}{r^3}ce^{-V}E,
\label{s2}
\ena
where we assume that $V$ does not depend on $r$.
Eliminating $r$-derivative terms
in Eq. \p{MOU32}, by Eqs. (\ref{s1}) and (\ref{s2}),
we obtain
\bea
\label{doream}
&&e^{2v-2u}\big[
\ddot{v}+\dot{v}\big(\dot{V}-2\dot{u}+2\dot{v}\big)
-v_{,yy}-v_{,y}
\big(V_{,y}-2u_{,y}+2v_{,y}\big)
\big]
-\frac{g^2}{\kappa^2}e^{\phi+2v}
\nonumber\\
&&=-\frac{1}{r^3}\Big(\tilde E_{,r}-\frac{2\tilde E^2}{r^3}\Big)
\ena
where we defined $\tilde E:=\kappa^2 ce^{-V}E$.
Setting the right-hand side of Eq. (\ref{doream}) to be zero,
we obtain
\bea
\tilde E=\frac{Q}{H(t,y,r)}, \quad
H(t,y,r) \equiv h(t,y)+\frac{Q}{r^2},
\ena
where $Q$ is a constant.
Then Eqs.~\p{s1} and \p{s2} give that
the metric functions and dilaton can be written as
\bea
u=-\frac14\ln H(t,y,r), \quad
v=\frac14\ln H(t,y,r), \quad
\phi = -\frac12\ln H(t,y,r),
\ena
where the integration constants are set to be zero.
It is straightforward
to check that the off-diagonal components of the Einstein equation
\bea
\label{MOU7}
&&
4u_{,r}\dot v-\dot u{}_{,r}
-3\dot v{}_{,r}
-\phi_{,r}\dot\phi=0,
\nonumber \\
&&
4u_{,r}v_{,y}-u_{,ry}
-3v_{,ry}
-\phi_{,r}\phi_{,y}=0,
\eea
are satisfied.
We find that the time dependent parts of
Eqs. (\ref{MOU12}), (\ref{MOU22}) and (\ref{MOU32}) reduce
to
\bea
h_{,yy}+3\ddot{h}=\frac{4g^2}{\kappa^2},
\label{red1}
\\
-\ddot{h}-3h_{,yy}=\frac{4g^2}{\kappa^2},
\label{red2}
\\
\ddot{h}-h_{,yy}=\frac{4g^2}{\kappa^2},
\label{red3}
\eea
respectively.
Both of Eqs. (\ref{MOU42}) and (\ref{MOU52})
reduce to the same equation as Eq. (\ref{red3}).
The solution of Eq. (\ref{red1})-(\ref{red3}) is given
by
\bea
\label{sol_h}
h(t,y)=\frac{g^2}{\kappa^2}\big(t^2-y^2\big)
+a_t t+a_y y+a_0
\eea
where $a_i$ ($i=t,y,0$) are integration constants.
Then, the left-hand side of the remaining off-diagonal component
of the Einstein equation
\bea
-4v_{,y}\dot{v}-4\dot{v}_{,y}+4u_{,y}\dot{v}
+4\dot{u}v_{,y}-\dot{\phi}\phi_{,y}=0,
\eea
is proportional to $\dot{h}_{,y}$, which
vanishes because of Eq. (\ref{sol_h}).
As seen from Eq. (\ref{sol_h}), $h$ is no longer linear in the worldvolume
coordinates but quadratic in them, which also happens in $p$-brane
solutions with trivial or vanishing dilaton \cite{KU}.

Let us summarize the solutions obtained here:
\bea
&&u(t,y,r)=-v(t,y,r)=-\frac{1}{4}\ln
\Big(\frac{Q}{r^2}+\frac{g^2}{\kappa^2}(t^2-y^2)
+a_t t+a_y y+a_0\Big),
\\
&&
\phi(t,y,r)=-\frac{1}{2}\ln
\Big(\frac{Q}{r^2}+\frac{g^2}{\kappa^2}(t^2-y^2)
+a_t t+a_y y+a_0\Big).
\eea
Because of the dependence on the spatial worldvolume coordinate $y$
as well as on the time coordinate $t$, the brane direction is not compact and
this solution may not be suitable for constructing the brane world.
One might imagine that there may be a solution with a quadratic order
dependence on time in the case of the 2-form field (i.e. 0-brane).
For the dilaton coupling parameters in the NSS model,
such a dynamical solution does not exist.
But it does for the other special coupling parameters.
Our solution can be interpreted as one of
the dynamical $p$-brane solutions,
in a class of the theory with a cosmological constant.
These solutions will be discussed in detail in \cite{ku}.
In particular, among these solutions,
the $0$-brane solution in the four-dimensional
spacetime behaves as a black hole embedded into
a FRW Universe filled by the matter with the equation of
state $w=-\frac{1}{3}$, where $w$ is the ratio of the
pressure with the energy density.

\end{document}